\documentclass[aps,twocolumn,amssymb,superscriptaddress]{revtex4-1}
\usepackage[final]{changes}
\usepackage[colorlinks=true,citecolor=blue,linkcolor=blue,urlcolor=blue]{hyperref}
\usepackage{amsmath}
\usepackage{amssymb}
\usepackage{bbm}
\usepackage{xcolor}
\usepackage{graphicx}
\usepackage{dcolumn}
\usepackage{bm}
\usepackage{hyperref}
\usepackage{lipsum}
\usepackage{verbatim}
\graphicspath{{Latexpicture/}}
\begin{document}
	\title{Ultrafast and deterministic generation of Bell states in the ultrastrong coupling regime}
\author{Xin Xie}
\affiliation{Hunan Key Laboratory for Micro-Nano Energy Materials and Devices\\ and School of
Physics and Optoelectronics, Xiangtan University, Hunan 411105, China}
\author{Junlong Tian}
\affiliation{Department of electronic science, College of Big Data and Information Engineering, Guizhou University, Guiyang 550025, China}
\author{Jie Peng}
\email{jpeng@xtu.edu.cn}
\affiliation{Hunan Key Laboratory for Micro-Nano Energy Materials and Devices\\ and School of
Physics and Optoelectronics, Xiangtan University, Hunan 411105, China}

\begin{abstract}
We have found  the special dark state solutions of the anisotropic two-qubit quantum Rabi model (QRM), which has at most one photon, and constant eigenenergy in the whole coupling regime. Accordingly, we propose a scheme to deterministically generate two kinds of the two-qubit Bell states through adiabatic evolution along the dark states. With the assistance of the Stark shift, the generation time can be reduced to subnanosecond scales,  proportional to the reverse of the resonator frequency, with fidelity reaching 99\%. Furthermore, the other two kinds of Bell states can also be ultrafast generated.
\end{abstract}
\maketitle
\emph{Introduction.}-The Rabi model, originally introduced by  Isidor Rabi, provides a fundamental framework for understanding the interaction between an electromagnetic field and a two-level quantum system \cite{Rabi}. In most cases, the coupling strength $g$ is much smaller than the photon frequency $\omega$, such that the counterrotating terms can be neglected, and the Jaynes-Cummings model \cite{JC} is widely applied in cavity and circuit QED \cite{Cqe,Mqe}. However, ultrastrong \cite{Ultra} ($g>0.1\omega$) and even deep strong couplings \cite{Deep} ($g>\omega$) have been realized in experiment recently, so it is important to study the QRM to explorer the application of these stronger couplings, which seems obviously, could accelerate the qubit-photon interaction. The problem is that the addition of counter-rotating terms will connect all the photon number states, such that the solutions to QRM generally involve subspaces with infinite photons \cite{braak,chen,Sedov,zhong,xie,yu}, making its dynamics quite complex and applications difficult in quantum computation. \par

While the standard Rabi model assumes symmetric coupling between the field and the system, many real-world scenarios exhibit anisotropic behaviors, where the coupling strength varies in different directions or under specific conditions  \cite{Xiang}. 
To address these cases, the anisotropic Rabi model \cite{Anir} has been developed, offering deeper insights and more accurate representations of physical systems.

Recently, the anisotropic Rabi model is extensively studied \cite{zhangyuyu1,zheng,zhangyuyu2,wang yimin,Shuai Cui,Sor,Xunda Jiang} and some interesting schemes are proposed to realize it \cite{zhihai,Amf}.  Understanding the implications of anisotropy on the Rabi dynamics \cite{Asyrs,Us} enables us to better comprehend and manipulate the behavior of such quantum systems. However, its solutions still involve subspaces with an unbounded number of photons \cite{Anir}, making its dynamics complex and blocking its way to applications in quantum information.\par

Here, we have found the one-photon special dark-state solutions to the anisotropic two-qubit QRM, which has at most one photon and constant eigenenergy in the whole coupling regime. The coherent superposition of basis with one photon will cancel higher photon excitation when applied by the Hamiltonian. These solutions can be used to deterministically generate two kinds of two-qubit Bell states  $(|\uparrow\downarrow\rangle+|\downarrow\uparrow\rangle)/\sqrt{2}$ and  $(|\downarrow\uparrow\rangle-|\uparrow\downarrow\rangle)/\sqrt{2}$ \cite{Bs,Qtele,Clb} through adiabatic evolution \cite{Qae,Neqs,Fae}. Interestingly, this adiabatic evolution can be quite fast due to the peculiarity of the dark states and reaching of the ultrastrong coupling regime. For a simple linear adiabatic trajectory, these Bell states can be generated with fidelity reaching $99\%$ in $49\omega^{-1}$, corresponding to $2.6$ ns if the resonator frequency $\omega=2\pi\times3$GHz. The generation time can be reduced to 34 $\omega^{-1}$ ($1.8$ns) if we choose a nonlinear adiabatic trajectory. Interestingly, the addition of the Stark shift to the two-qubit anisotropic QRM \cite{Asyrs} will reduce the generation time to $9.8\omega^{-1}$ ($0.52$ns) even for a simple linear adiabatic trajectory. This time is proportional to the reverse of the resonator frequency $\omega/2\pi$, showing a sign of ultrafast state-generation \cite{ultrafast}.  Furthermore, other two kinds of two-qubit Bell states $(|\downarrow\downarrow\rangle+|\uparrow\uparrow\rangle)/\sqrt{2}$ and  $(|\downarrow\downarrow\rangle-|\uparrow\uparrow\rangle)/\sqrt{2}$  can also be ultrafast generated in $9\omega^{-1}$ ($0.48$ ns). 

 \par
\emph{One-photon dark state solutions to the anisotropic two-qubit QRM.}--
The anisotropic two-qubit   QRM  reads\cite{Anir}
\begin{equation}\label{H}
\begin{aligned}
	H=&\omega a^\dagger a +\sum_j\Delta_j\sigma_{jz}
	+\sum_j[g_{jr}(a^\dagger \sigma_{j}+a\sigma_j^\dagger)\\
	&+g_{jc}(a^\dagger\sigma_j^\dagger+a\sigma_j)]
\end{aligned} 
\end{equation}
where $a^\dagger$ is the photonic creation operator and $a$ is the annihilation operator with frequency $\omega$. $\sigma_{jz}$ and $\sigma_{jx}$ are Pauli operators for the $j$-th qubit with frequency $2\Delta_j$ ($j=1,2$). $g_{jr}$ and $g_{jc}$ denote the coupling strength of the rotating terms and counter-rotating terms respectively. 
 Hamiltonian Eq. \eqref{H} has a $\mathbb{Z}_2$ symmetry \cite{braak,AG,QRN,Nsrm,QRM} with operator $R=e^{i\pi a^\dagger a} \sigma_{1z} \sigma_{2z}$, which can be used to classify basis by even parity and odd parity. However, $H$ is still infinite dimensional in the parity subspace, and general solutions consist of infinite photons. 

However, although the interaction terms connect all the Fock states, it is possible to find some quasi-exact solutions with finite photon number states, because the coherent superposition of them could cancel the possible population of higher photon number states when applied by $H$ \cite{Dsm,Os,As}.   

Supposing there are eigenstates with at most one photon $|\Psi_1\rangle=c_1|0\uparrow\uparrow\rangle+c_2|0\downarrow\downarrow\rangle+c_3|1\uparrow\downarrow\rangle+c_4|1\downarrow\uparrow\rangle$ and $|\Psi_2\rangle=c_1|0\uparrow\downarrow\rangle+c_2|0\downarrow\uparrow\rangle+c_3|1\uparrow\uparrow\rangle+c_4|1\downarrow\downarrow\rangle$ with even and odd parity respectively, the eigenenergy equations read
\begin{widetext}
	\begin{equation}\label{Even Matrix}
		\left(
		\begin{array}{cccc}
			\Delta_1+\Delta_2-E & 0 & g_{2r} & g_{1r}\\
			0 & -\Delta_1-\Delta_2-E & g_{1c} & g_{2c}\\
			g_{2r} & g_{1c} & \omega+\Delta_1-\Delta_2-E & 0\\
			g_{1r} & g_{2c} & 0 & \omega-\Delta_1+\Delta_2-E\\
			0 & 0 & \sqrt{2}g_{2c} & \sqrt{2}g_{1c}\\
			0 & 0 & \sqrt{2}g_{1r} & \sqrt{2}g_{2r}\\
		\end{array} 
		\right)
		\left(
		\begin{array}{c}
			c1\\c2\\c3\\c4\\
		\end{array}\right)=0,
	\end{equation} 
	\begin{equation}\label{Odd Matrix}
		\left(
		\begin{array}{cccc}
			\Delta_1-\Delta_2-E & 0 & g_{2c} & g_{1r}\\
			0 & -\Delta_1+\Delta_2-E & g_{1c} & g_{2r}\\
			g_{2c} & g_{1c} & \omega+\Delta_1+\Delta_2-E & 0\\
			g_{1r} & g_{2r} & 0 & \omega-\Delta_1-\Delta_2-E\\
			0 & 0 & \sqrt{2}g_{2r} & \sqrt{2}g_{1c}\\
			0 & 0 & \sqrt{2}g_{1r} & \sqrt{2}g_{2c}\\
		\end{array} 
		\right)
		\left(
		\begin{array}{c}
			c1\\c2\\c3\\c4\\
		\end{array}\right)=0,
	\end{equation}\\
\end{widetext}
respectively.

\begin{figure}[tbhp]
	\centering 
	\resizebox{1\columnwidth}{!}{
		\includegraphics{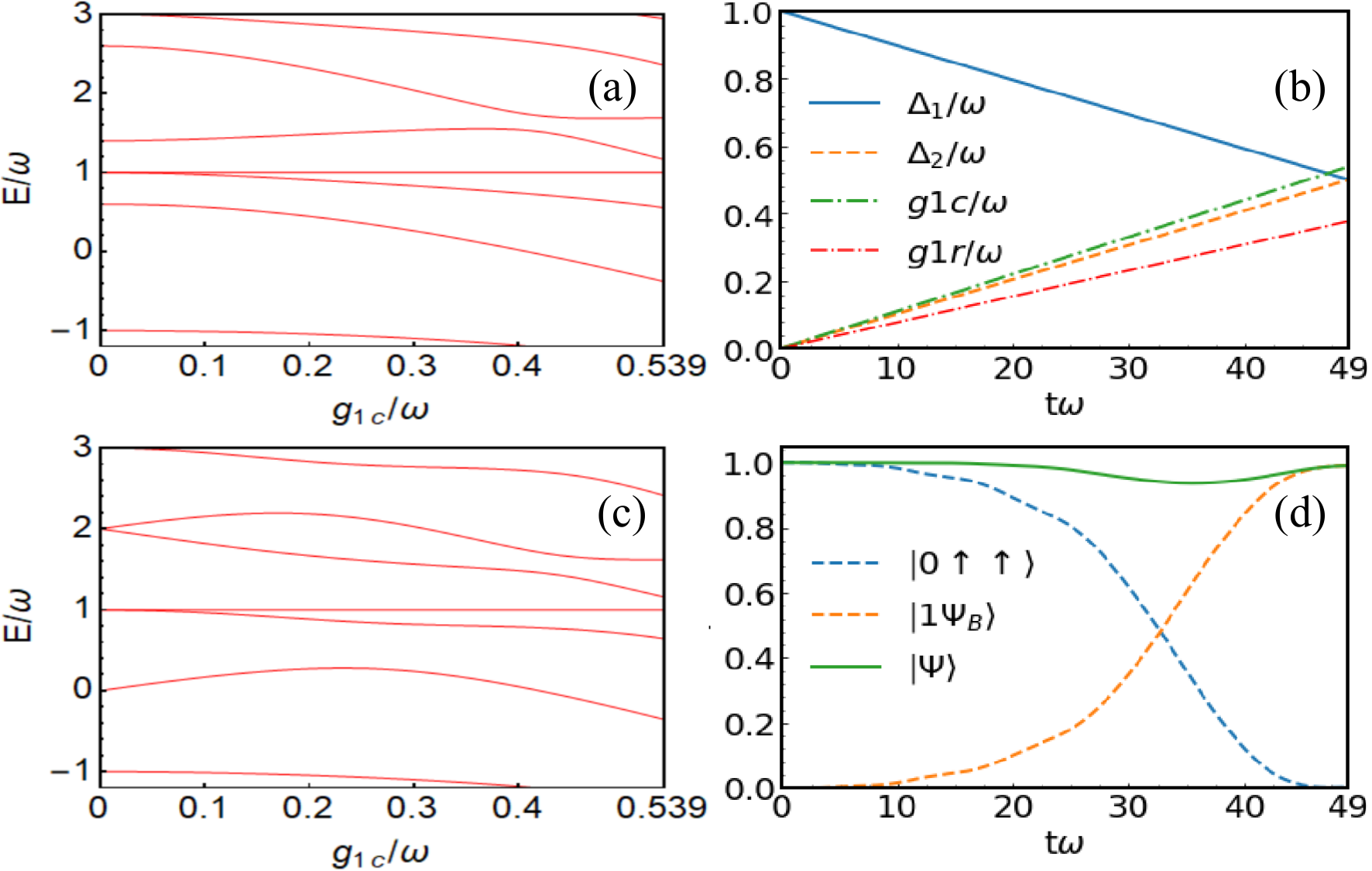}
	}
	\caption[1]{(a) Even-parity spectrum of the anisotropic two-qubit QRM. $\Delta_1 =0.7\omega$,~$\Delta_2=0.3\omega$, $g_{1r}=0.7g_{1c}$,~$g_{2r}=0.7g_{2c}$, $g_{1c}=g_{2c}$.  (b) The evolution of parameter in adiabatic evolution to obtain $|\psi_B\rangle=\frac{1}{\sqrt{2}}(-|1\uparrow\downarrow\rangle+|1\downarrow\uparrow\rangle)$ with  $g_{1r}=0.7g_{2c}$, $g_{2r}=0.7g_{1c}$, $g_{1c},~g_{2c} $grow linearly from 0 to 0.539$\omega$.  $\Delta_1,\Delta_2$ grow from $\omega$ and $0$ to $\omega/2$ linearly, respectively.  (c) Spectrum of the anisotropic two-qubit QRM with parameters shown in (b). (d) The population of states when parameters evolve as (b) in the adiabatic evolution. } 
	\label{fig1}  
\end{figure}

If there is less row number than column number in the above matrices after elementary row transformations, then there are nontrivial solutions \cite{Os}. This can be done when $\omega=\Delta_1+\Delta_2=E$, $g_{1r}/g_{2r}=g_{1c}/g_{2c}=\pm1$ for even parity,
and the matrix become 
\begin{equation}
	\begin{aligned}
		&\left(
		\begin{array}{cccc}
			g_{1r}&0&0&-\Delta_1+\Delta_2\\
			0&1&0&0\\
			0&0&1&\pm1\\
			0&0&0&0\\
			0&0&0&0\\
			0&0&0&0\\
		\end{array} 
		\right)	
	\end{aligned}.
\end{equation}\\
The corresponding  eigenstates read

\begin{equation}
	|\Psi\rangle=\frac{1}{\cal{N}}[(\Delta_1-\Delta_2)|0\uparrow\uparrow\rangle+g_{1r}\vert 1\rangle(\vert\downarrow\uparrow\rangle\mp|\uparrow\downarrow\rangle)].\label{state1}
\end{equation}
For any available coupling values, the eigenenergy is a constant $E=\omega$, corresponding to a horizontal line in the spectrum, as shown in Fig. \ref{fig1} (a).


For odd parity, there are special dark states
\begin{equation}
	|\Psi\rangle=\frac{1}{\cal{N}}[(\Delta_1+\Delta_2)|0\uparrow\downarrow\rangle+g_{1r}|1\rangle(\vert\downarrow\downarrow\rangle\mp\vert\uparrow\uparrow\rangle)]\label{state3}
\end{equation}
with constant eigenenergy $E=\omega$ under the conditions $\Delta_1-\Delta_2=\omega$, $g_{2r}/g_{1c}= g_{1r}/g_{2c}=\pm1$, and special dark state
\begin{equation}
	|\Psi\rangle=\frac{1}{\cal{N}}[(\Delta_1+\Delta_2)|0\downarrow\uparrow\rangle+g_{2r}|1\rangle(\vert\downarrow\downarrow\rangle\mp\vert\uparrow\uparrow\rangle)]\label{state5}
\end{equation}
with constant eigenenergy $E=\omega$ under the condition $\Delta_2-\Delta_1=\omega$, $g_{2r}/g_{1c}= g_{1r}/g_{2c}=\pm1$.

\emph{Fast and deterministic generation of Bell states.}-Dark states Eqs. \eqref{state1}-\eqref{state3} consist of four kinds of Bell states $(\vert\downarrow\uparrow\rangle\mp|\uparrow\downarrow\rangle)/\sqrt{2}$, $(\vert\downarrow\downarrow\rangle\mp|\uparrow\uparrow\rangle)/\sqrt{2}$. So it is a natural consideration to generate them through adiabatic evolution along these dark states by tuning the ratio between $\Delta_1\mp\Delta_2$ and $g_{jr}$. However, it is impossible to make $g_{jr}\gg\Delta_1+\Delta_2$ under the condition $|\Delta_1-\Delta_2|=\omega$ in current experiments, so we first focus on the generation  of $(\vert\downarrow\uparrow\rangle\mp|\uparrow\downarrow\rangle)/\sqrt{2}$.


Our scheme is as follows: First we cool down the two qubits with $\Delta_1\neq\Delta_2$ and a resonator to ground state $\vert 0\downarrow\downarrow\rangle$, and then we excite the two qubits to obtain $\vert 0\uparrow\uparrow\rangle$. The qubit and resonator are initially decoupled, so $\vert 0\uparrow\uparrow\rangle$ just corresponds to dark state Eq. \eqref{state1}. Then we decrease $\Delta_1\neq\Delta_2$ to zero and increase $g_{1r}$ to a nonzero value, such that $(\vert\downarrow\uparrow\rangle\mp|\uparrow\downarrow\rangle)/\sqrt{2}$ are generated at the end of the adiabatic evolution. When parameters evolves as Fig. \ref{fig1} (b), $\vert\psi_B\rangle=(\vert\downarrow\uparrow\rangle-|\uparrow\downarrow\rangle)/\sqrt{2}$ can be generated in $49\omega^{-1}$ with fidelity reaching $99\%$, as shown in Fig. \ref{fig1} (d), which is faster than isotropic case \cite{Os}. If we choose $g_{2r}/g_{1r}=g_{2c}/g_{1c}=-1$, then $(\vert\downarrow\uparrow\rangle+|\uparrow\downarrow\rangle)/\sqrt{2}$ can be generated with the same speed and fidelity.

The spectrum during the adiabatic evolution is shown in Fig. \ref{fig1} (c). There   is an energy level $\vert\psi\rangle$ very close to the dark state $\vert \Psi\rangle$, but surprisingly, the adiabatic evolution along $\vert\Psi\rangle$ is successful and fast, which is counterintuitive according to the adiabatic theorem \cite{Cat,Ep}, which states that a system will evolve adiabatically along eigenstate $\vert E_n(t)\rangle$ if 
\begin{equation}
	\bigg\lvert\frac{\langle E_m(t)|\dot{H}|E_n(t)\rangle}{(E_m-E_n)^2}\bigg\rvert \ll 1, \quad
	t \in [0,T]
	\end{equation} 
Since $\vert\psi_{E=\omega+\delta}\rangle$ is very close to $\vert \Psi_{E=\omega}\rangle$, normally $\dot{H}$ has to be very small.
However, because the peculiarity of $\vert \Psi\rangle$, $\langle \psi_{E=\omega+\delta}|\dot{H}|\Psi(t)\rangle\propto f(\Delta_{1,2},g)|\delta|/\sqrt{(\Delta_1-\Delta_2)^2+\delta^2}$, which decrease as $\vert \psi_{E=\omega+\delta}\rangle$ goes closer to $\vert \Psi\rangle$,   ensuring the successful adiabatic evolution. Taking the adiabatic evolution along dark state Eq. \eqref{state1} for example, 
\begin{align}
	\dot{H}=&\dot{\Delta}_1\sigma_{z1}-\dot{\Delta}_1\sigma_{z2}+\dot{g}_{1r}(a^\dagger \sigma_{1}+a\sigma_1^\dagger)\notag \\
	\pm&\dot{g}_{1r}(a^\dagger\sigma_2+a\sigma_2^\dagger)
	+\dot{g}_{1c}(a^\dagger\sigma_1^\dagger+a\sigma_1)\notag\\
	&\pm\dot{g}_{1c}(a\sigma_2+a^\dagger\sigma_2^\dagger),\notag \\
\end{align}
where $\dot{\Delta}_1=-\dot{\Delta}_2$ to ensure $\Delta_1(t)+\Delta_2(t)=\omega$, such that 
\begin{equation}\label{co}
\dot{H}|\Psi\rangle=\frac{1}{{\cal
N}}|1\rangle(|\downarrow\uparrow\rangle\pm
 |\uparrow\downarrow\rangle)\left[\dot{g}_{1r}(\Delta_1-\Delta_2)-2\dot{\Delta}_1 g_{1r}
\right].
\end{equation}
It can be seen that only $\vert1\uparrow\downarrow\rangle$ and $\vert1\downarrow\uparrow\rangle$ in $\vert E_m\rangle$ will affect the adiabatic speed along $\vert \Psi\rangle$.
Projecting $(H-\omega-\delta)|\psi_{E_m=\omega+\delta}\rangle=0$ into $\vert 1\downarrow\uparrow\rangle-\vert 1\uparrow\downarrow\rangle$, we arrive at
\begin{eqnarray}
(\Delta_2-\Delta_1-\delta)\langle 1\downarrow\uparrow|\psi_{E_m=\omega+\delta}\rangle\nonumber\\=(\Delta_1-\Delta_2-\delta)\langle 1\uparrow\downarrow|\psi_{E_m=\omega+\delta}\rangle.
\end{eqnarray}
Therefore,
\begin{equation}
\label{coe}
\langle\psi_{E=\omega+\delta}|\dot{H}|\Psi\rangle\propto f(\Delta_{1,2},g)|\delta|/\sqrt{(\Delta_1-\Delta_2)^2+\delta^2},
\end{equation}
considering Eq. \eqref{co}.
The same results can be obtained for the other dark states Eqs. \eqref{state3} and \eqref{state5}.

\begin{figure}[tbhp]
	\centering
	\includegraphics[scale=0.45]{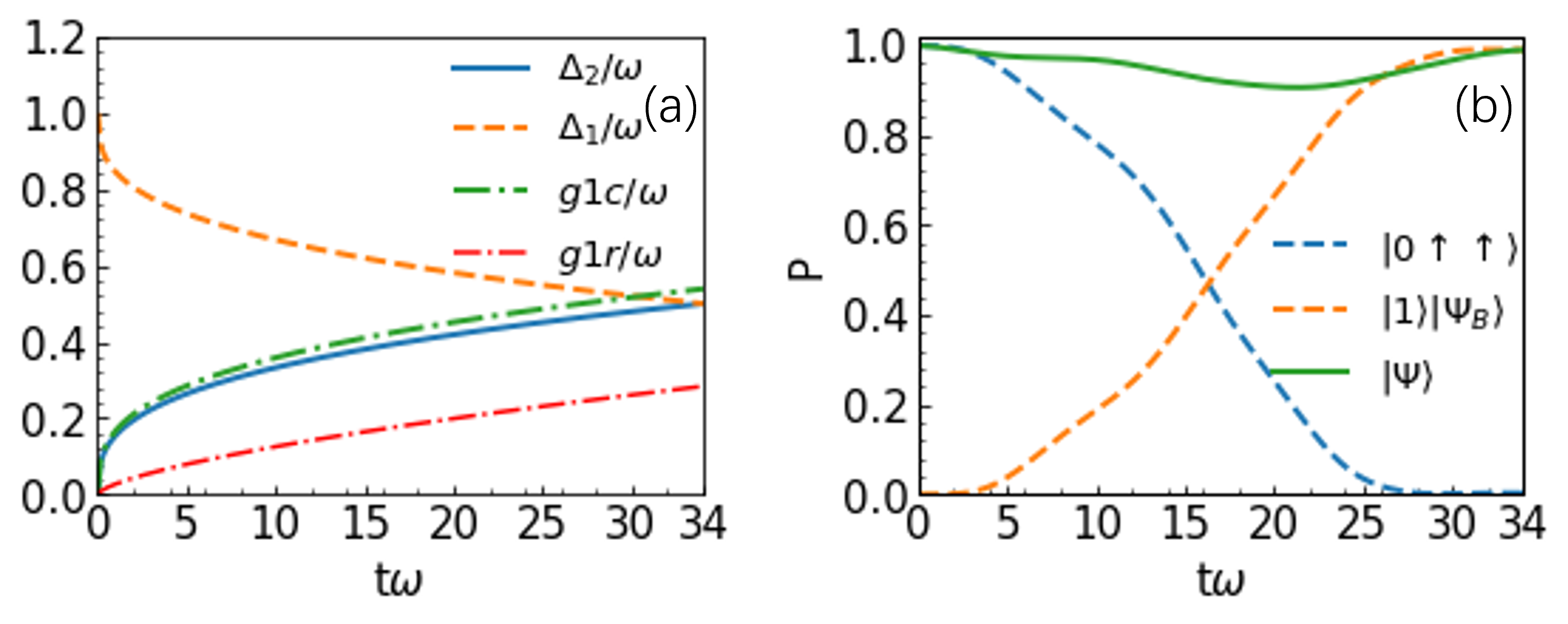}
	\renewcommand\figurename{\textbf{FIG.}}
	\caption[4]{(a) The nonlinear trajectory of parameters to generate $\vert \psi_B\rangle$ through adiabatic evolution along dark state Eq. (\ref{state1}). $\Delta_1/\omega$ =$1-\frac{1}{2} (\frac{\omega t}{34})^\frac{1}{3}$. $\Delta_2/\omega$ =$\frac{1}{2} (\frac{\omega t}{34})^\frac{1}{3}$ and $g_{1c}=g_{2c} =0.539(\frac{\omega t}{34})^\frac{1}{3}$. $g_{1r}=g_{2r}=0.284(\frac{\omega t}{34})^\frac{2}{3}$. (b) The population of states during the adiabatic evolution.}
	\label{fig2}
\end{figure}
\begin{figure}[bthp]
	\centering
	\includegraphics[scale=0.5]{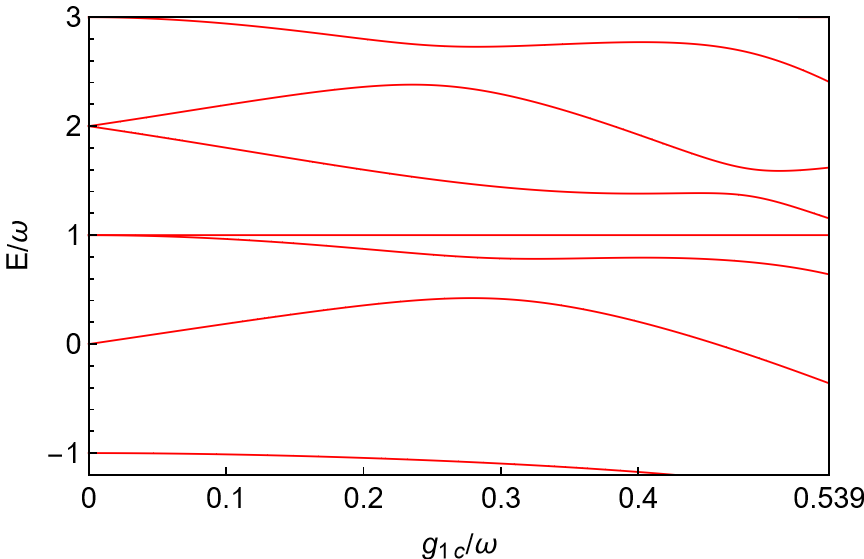}
	\renewcommand\figurename{\textbf{FIG.}}
	\caption[3]{Even-parity spectrum of the anisotropic two-qubit QRM during the adiabatic evolution shown in Fig. \ref{fig3}. }
	\label{fig3}
\end{figure}

To further accelerate the adiabatic evolution, we use a nonlinear adiabatic trajectory as shown in Fig. \ref{fig2} (a), where $\vert \psi_B\rangle$ can be generated in $34\omega^{-1}$ with fidelity still reaching 99\%, as shown in \ref{fig2} (b). The corresponding spectrum is depicted in   Fig. \ref{fig3}.

\begin{figure}[b]
	\centering
	\includegraphics[scale=0.55]{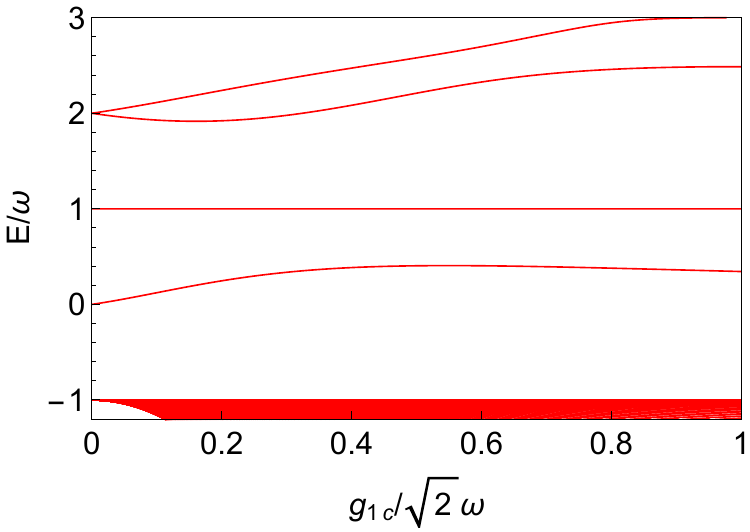}
	\renewcommand\figurename{\textbf{FIG.}}
	\caption[5]{Even-parity spectrum of the two-qubit Rabi-Stark model, where $g_{1r}=g_{2r},g_{1c}=g_{2c},g_{1r}=0.98g_{1c}$. $\Delta_2=g_{1c}/2\sqrt{2}$. $\Delta_1=\omega-\Delta_2$. $U_1=U_2=\omega/2$.}
	\label{fig4}
\end{figure}

\emph{Ultrafast generation of all Bell states.}--There are still two tasks for us to accomplish: The first one is to speed up the adiabatic evolution to realize ultrafast Bell-states generation, and the second one is to  generate the rest two kinds of Bell states, $(|\downarrow\downarrow\rangle\mp|\uparrow\uparrow)/\sqrt{2}$. It can be seen from spectrum Figs. \ref{fig1} and \ref{fig4} that $\vert\Psi\rangle$ Eq. \eqref{state1} and $\vert 2\downarrow\downarrow\rangle$ are degenerate at $g=0$, such that there is always an energy level very close to $\vert \Psi\rangle$ when $g_{1r}$ is small. Intuitively, they can be separated by the Stark shift terms $a^\dagger a\sigma_{z}$, so we consider the anisotropic two-qubit Rabi-Stark model \cite{Asyrs}
\begin{equation}\label{Hs}
\begin{aligned}
    H={}&(\omega+\sum_j U_j\sigma_{jz}) a^\dagger a+\sum_j\Delta_j\sigma_{jz}\\
    &+\sum_j[g_{jr}(a^\dagger \sigma_{j}+a\sigma_j^\dagger)
	+g_{jc}(a^\dagger\sigma_j^\dagger+a\sigma_j)],
\end{aligned}
\end{equation}
as proposed in \cite{Smu} to ultrafast generate the single-photon multimode $W$ states.

This Hamiltonian still possesses the $\mathbb{Z}_2$ symmetry with operator $R=e^{i\pi a^\dagger a} \sigma_{1z} \sigma_{2z}$ \cite{braak,AG,QRN,Nsrm,QRM}, and is only well-defined when $|U_1+U_2|\leq\omega$ \cite{Smu,Criti}, because or else its energies has no lower bound, as can be seen in Eq. \eqref{Hs}. Supposing there are eigenstates with at most one photon $|\Psi_1\rangle=c_1|0\uparrow\uparrow\rangle+c_2|0\downarrow\downarrow\rangle+c_3|1\uparrow\downarrow\rangle+c_4|1\downarrow\uparrow\rangle$ and $|\Psi_2\rangle=c_1|0\uparrow\downarrow\rangle+c_2|0\downarrow\uparrow\rangle+c_3|1\uparrow\uparrow\rangle+c_4|1\downarrow\downarrow\rangle$ with even and odd parity respectively, the eigenenergy equations read
\begin{widetext}
	\begin{equation}
			\left(
		\begin{array}{cccc}
			\Delta_1+\Delta_2-E & 0 & g_{2r} & g_{1r}\\
			0 & -\Delta_1-\Delta_2-E & g_{1c} & g_{2c}\\
			g_{2r} & g_{1c} & \omega+\Delta_1-\Delta_2+u_1-u_2-E & 0\\
			g_{1r} & g_{2c} & 0 & \omega-\Delta_1+\Delta_2-u_1+u_2-E\\
			0 & 0 & \sqrt{2}g_{2c} & \sqrt{2}g_{1c}\\
			0 & 0 & \sqrt{2}g_{1r} & \sqrt{2}g_{2r}\\
		\end{array} 
		\right)
		\left(
		\begin{array}{c}
			c1\\c2\\c3\\c4\\
		\end{array}\right)=0
	\end{equation}
		\begin{equation}
			\left(
			\begin{array}{cccc}
				\Delta_1-\Delta_2-E & 0 & g_{2c} & g_{1r}\\
				0 & -\Delta_1+\Delta_2-E & g_{1c} & g_{2r}\\
				g_{2c} & g_{1c} & \omega+\Delta_1+\Delta_2+u_1+u_2-E & 0\\
				g_{1r} & g_{2r} & 0 & \omega-\Delta_1-\Delta_2-u_1-u_1-E\\
				0 & 0 & \sqrt{2}g_{2r} & \sqrt{2}g_{1c}\\
				0 & 0 & \sqrt{2}g_{1r} & \sqrt{2}g_{2c}\\
			\end{array} 
			\right)
			\left(
			\begin{array}{c}
				c1\\c2\\c3\\c4\\
			\end{array}\right)=0
		\end{equation}
\end{widetext}

For even parity, such solution reads
\begin{equation}
	|\Psi\rangle=\frac{1}{\cal{N}}[(\Delta_1-\Delta_2+U_1-U_2)|0\uparrow\uparrow\rangle+g_{1r}\vert 1\rangle(\vert\downarrow\uparrow\rangle\mp|\uparrow\downarrow\rangle)],\label{stark1}
\end{equation}
with constant eigenenergy in the whole coupling regime under the condition $\omega=\Delta_1+\Delta_2=E$, $g_{2r}/g_{1r}=g_{2c}/g_{1c}=\pm1$,
as shown in Fig. \ref{fig4}.
For odd parity, there are special dark states
\begin{equation}
	|\Psi\rangle=\frac{1}{\cal{N}}[(\Delta_1+\Delta_2+U_1+U_2)|0\uparrow\downarrow\rangle+g_{1r}|1\rangle(\vert\downarrow\downarrow\rangle\mp\vert\uparrow\uparrow\rangle)]\label{stark3}
\end{equation}
with constant eigenenergy $E=\omega$ under the conditions $\omega=\Delta_1-\Delta_2$,~$g_{2r}/g_{1c}= g_{1r}/g_{2c}=\pm 1$, and special dark state
\begin{equation}
	|\Psi\rangle=\frac{1}{\cal{N}}[(\Delta_1+\Delta_2+U_1+U_2)|0\downarrow\uparrow\rangle+g_{2r}|1\rangle(\vert\downarrow\downarrow\rangle\mp\vert\uparrow\uparrow\rangle)]\label{stark5}
\end{equation}
with constant eigenenergy $E=\omega$ under the condition $\omega=-\Delta_1+\Delta_2$, $g_{2r}/g_{1c}= g_{1r}/g_{2c}=\pm 1$.

\begin{figure}[htbp]
	\centering
	\includegraphics[scale=0.45]{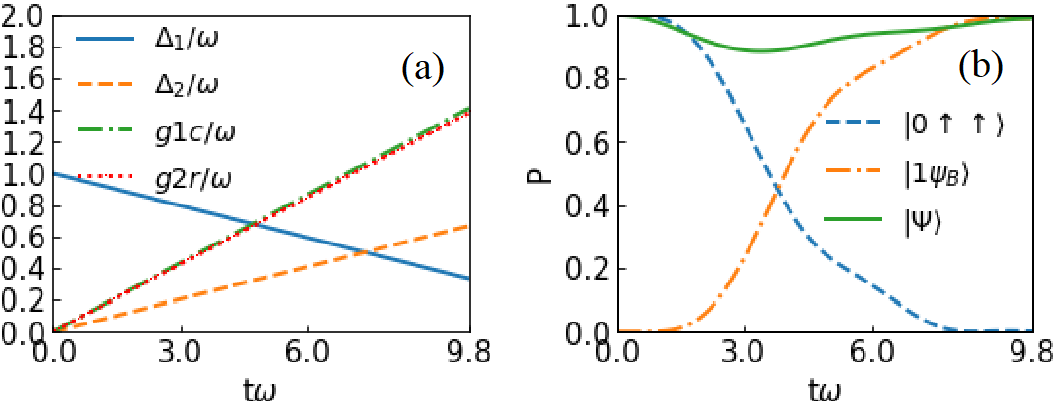}
	\renewcommand\figurename{\textbf{FIG.}}
	\caption[6]{(a) The change of parameters in the adiabatic evolution to obtain $\frac{1}{\sqrt{2}}(-|1\uparrow\downarrow\rangle+|1\downarrow\uparrow\rangle)$ through dark state Eq. \eqref{stark1}. $\Delta_1/\omega$ evolves from 1 to $\frac{1}{3}$. $\Delta_2/\omega$ evolves from 0 to $\frac{2}{3}$. $g_{1c}/\omega$ evolves from 0 to $\sqrt{2}$. $g_{1r}=0.98 g_{1c}$. $U_1/\omega=\frac{2}{3}$, $U_2/\omega=\frac{1}{3}$. (b)Population of states during this adiabatic evolution.}
	\label{fig5}
\end{figure}

Now our first task can be accomplished using dark state Eq. \eqref{stark1} since the energy gap between $\vert \Psi\rangle$ and other energy levels is enlarged. When parameters evolve as in Fig. \ref{fig5} (a), $\vert \psi_B\rangle=(\vert \downarrow\uparrow\rangle-\vert \uparrow\downarrow\rangle)/\sqrt{2}$ can be generated with fidelity over $99\%$ in $9.8\omega^{-1}$, corresponding to $0.52$ ns when $\omega=2\pi\times 3$GHz, showing a sign of ultrafast state-generation \cite{ultrafast}, as shown in Fig. \ref{fig5} (b). If we simply choose $g_{1r}=-g_{2r}$ and $g_{1c}=-g_{2c}$, then another Bell state $\vert \psi_B\rangle=(\vert \downarrow\uparrow\rangle+\vert \uparrow\downarrow\rangle)/\sqrt{2}$ can be generated with the same speed and fidelity.

\begin{figure}[tbhp]
	\centering
	\includegraphics[scale=0.43]{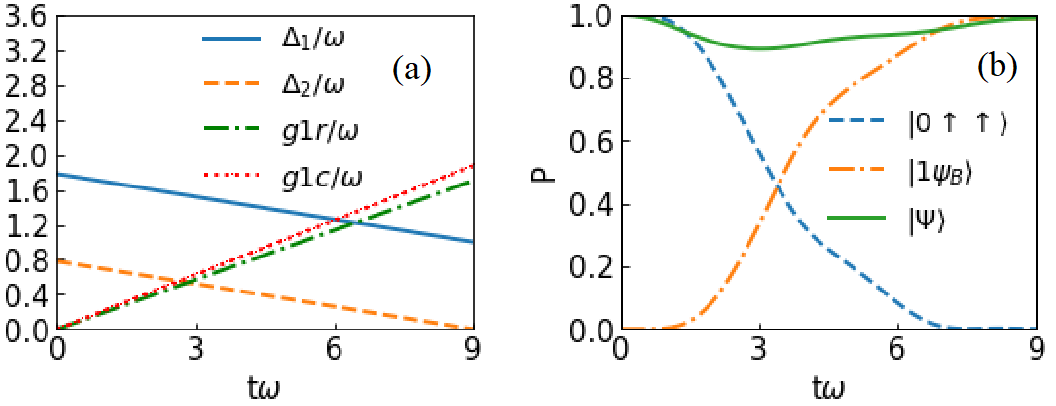}
	\renewcommand\figurename{\textbf{FIG.}}
	\caption[7]{(a)The change of parameters in the adiabatic evolution to obtain $\frac{1}{\sqrt{2}}(|1\downarrow\downarrow\rangle-|1\uparrow\uparrow\rangle)$ through dark state Eq. \eqref{stark3}. $\Delta_1$ evolves from $16\omega/9$ to $\omega$. $\Delta_2$ evolves from $7\omega/9$ to 0. $g_{1c}$ evolves from 0 to $1.87\omega$. $g_{1r}$ evolves from 0 to $1.7\omega$. $U_1=0$. $U_2=-\omega$. (b) Population of states during this adiabatic evolution.}
	\label{fig6}
\end{figure}

Our second task can be done with the dark state Eq. \eqref{stark3} or \eqref{stark5}. If parameters evolve as Fig. \ref{fig6} (a), then $(\vert\downarrow\downarrow\rangle-\vert \uparrow\uparrow\rangle)/\sqrt{2}$ can be generated with fidelity over $99\%$ in $9\omega^{-1}$ ($0.48$ns),  as shown in Fig. \ref{fig6} (b).  The last Bell state $(\vert\downarrow\downarrow\rangle+\vert \uparrow\uparrow\rangle)/\sqrt{2}$ can be also generated with the same speed by choosing $g_{1c}=-g_{2r}$ and $g_{2c}=-g_{1r}$.

\emph{Conclusion}-We have found special dark-state solutions to the anisotropic QRM with at most one photon and constant eigenenergy in the whole coupling regime. They consist of different Bell states, such that $(\vert \downarrow\uparrow\rangle\mp\vert \uparrow\downarrow)/\sqrt{2}$ can be generated through adiabatic evolution along them. Their peculiarities and reach of ultrastrong coupling regime make these adiabatic evolution quite fast. A simple linear adiabatic trajectory leads to generation time $49\omega^{-1}$ with fidelity reaching $99\%$, corresponding to $2.6$ns for $\omega=2\pi\times 3$GHz in circuit QED experiments. This time can be shortened to $34\omega^{-1}$ ($1.8$ns) for a nonlinear trajectory. When we further include the Stark shift, the generation time will be proportional to the reverse of the resonator frequency $9.8\omega^{-1}$ ($0.52$ns). Furthermore, the other two Bell states $(\vert \downarrow\downarrow\rangle\mp\vert \uparrow\uparrow)\sqrt{2}$ can be ultrafast generated in $9\omega^{-1}$ ($0.48$ns).


\end{document}